\documentclass{elsart}
\usepackage{epsfig}
\usepackage{epstopdf}

\usepackage{graphics,epsfig}
\usepackage{graphicx}
\usepackage{bm}
\usepackage{amsmath}
\usepackage{mathrsfs}
\usepackage[dvips]{color}

\newcommand{\be}{\begin{equation}}
\newcommand{\ee}{\end{equation}}
\newcommand{\beq}{\begin{eqnarray}}
\newcommand{\eeq}{\end{eqnarray}}

\def\nue{\mathrel{{\nu_e}}}
\def\numu{\mathrel{{\nu_\mu}}}
\def\nutau{\mathrel{{\nu_\tau}}}
\def\nux{\mathrel{{\nu_x}}}

\def\barnue{\mathrel{{\bar \nu}_e}}
\def\barnumu{\mathrel{{\bar \nu}_\mu}}
\def\barnutau{\mathrel{{\bar \nu}_\tau}}

\def \lta {\mathrel{\vcenter{\hbox{$<$}\nointerlineskip\hbox{$\sim$}}}}
\def \gta {\mathrel{\vcenter{\hbox{$>$}\nointerlineskip\hbox{$\sim$}}}}

\def\t13{\mathrel{{\theta_{13}}}}
\def\y12{\mathrel{{\tan^2 \theta_{12}}}}
\def\c2{\mathrel{{\chi^2 }}}

\newcommand{\df}{DSN$\nu$F}
\newcommand{\snr}{SNR}
\newcommand{\sfr}{SFR}

\newcommand{\sn}{supernova}
\newcommand{\sne}{supernovae}

\begin{document}
\begin{frontmatter}


\title{The diffuse supernova neutrino flux, supernova rate and SN1987A}

\author{Cecilia Lunardini}
 
\address{Institute for Nuclear Theory and University of Washington, Seattle, WA 98195 }%

                             
\begin{abstract}
I calculate the diffuse flux of electron antineutrinos from all
supernovae using the information on the neutrino spectrum from SN1987A
and the information on the rate of supernovae from direct supernova
observations.  The interval of flux allowed at 99\% confidence level
is $\sim 0.05 - 0.35 ~{\rm cm^{-2} s ^{-1}}$ above the SuperKamiokande
(SK) energy cut of 19.3 MeV.  This result is at least a factor of
$\sim 4$ smaller than the current SK upper limit of 1.2 ${\rm cm^{-2}
s ^{-1}}$, thus motivating the experimental efforts to lower the
detection energy threshold or to upgrade to higher volumes. A Megaton water
Cherenkov detector with $\sim 90\%$ efficiency would record $\sim 2 -
44 $ inverse beta decay events a year depending on the energy cut.
\end{abstract}

\end{frontmatter}

\section{Introduction}  
\label{intro}
 
Neutrinos from core collapse supernovae are unique messengers of
information on the physics of supernovae and on the properties of
neutrinos.  About the former, neutrinos are precious to study events
that occur near the core of the star, where matter is opaque to
photons: the neutronization due to electron capture, the infall phase,
the formation and propagation of the shockwave and the cooling
phase. Moreover, they allow to test the cosmological rate of supernova
neutrino bursts and thus to probe indirectly the history of star
formation.  Within neutrino physics, one can learn about the hierarchy
(ordering) of the neutrino mass spectrum, about the e-3 entry of the neutrino
mixing matrix, about possible non-standard neutrino interactions,
existence of sterile states, etc..

The experimental study of supernova neutrinos is challenging in many
respects. With current and upcoming neutrino telescopes, two scenarios
are possible.  The first is the detection of a burst from an
individual supernova which is close enough to us to produce a
significant number of events in a detector. This limits the candidate stars to those
within few hundreds of kiloparsecs from the Earth, where the rate of
core collapse is as low as $\sim 2 - 3$ per century (see
e.g. \cite{Arnaud:2003zr,Ando:2005ka}). This explains why only one neutrino
signal of this type has been recorded so far, in 1987 from SN1987A
\cite{Hirata:1987hu,Hirata:1988ad,Bionta:1987qt}.  
The second
possibility is to study the diffuse flux of neutrinos from all
supernovae. This requires massive detectors with a difficult background rejection \cite{Malek:2002ns,Eguchi:2003gg} or very precise geochemical tests \cite{Haxton:1987bf,Bahcallcomment}.
So far, the searches for this flux have given negative
results, and upper limits were put. Among them, the most stringent is
given by SuperKamiokande (SK) \cite{Malek:2002ns}  on the flux of electron
antineutrinos above $E=19.3$ MeV in neutrino energy (see also the looser
constraint from KamLAND in the energy interval $8.3 - 14.8$ MeV \cite{Eguchi:2003gg}). This limit, at
90\% confidence level, is: 
\be \Phi(E>19.3~{\rm MeV})<1.2 ~{\rm
cm^{-2} s^{-1}}~,
\label{sklim}
\ee and holds for a variety of theoretical inputs
\cite{Malek:2002ns}\footnote{The bound in Eq. (\ref{sklim}) depends on
the energy spectrum of the neutrinos arriving at Earth.  Thus it is
model-dependent, even though different models of spectra happen to
give similar values of it \cite{Malek:2002ns}. }.  The bound
(\ref{sklim}) approaches the range of theoretical predictions, and
thus motivates the expectation that a positive signal may be seen in
the near future, either with more statistics at SK or at the next
generation Cherenkov detectors with Megaton volumes (20 times larger
than SK) like UNO \cite{Jung:1999jq,Wilkes:2005rg}, HyperKamiokande
(HK) \cite{Nakamura:2003hk}, and MEMPHYS \cite{Mosca:2005mi}.

When assessing the possibility of a future measurement, one should
 consider that predictions of the diffuse supernova neutrino flux
 (\df) suffer large uncertainties, due to our poor knowledge of the
 underlying physics. In particular, there are two sources of
 theoretical error. One is the uncertain value of the supernova
 rate  \footnote{In this context  \sn\ rate means the rate of neutrino-emitting supernovae. It does not include objects that do not produce neutrinos in significant amount, such as the type Ia supernovae or core collapse \sne\ that evolve into black holes before emitting neutrinos.}
(\snr\ for
 brevity) as a function of the redshift.  This rate can be obtained
 directly from supernova observations \cite{cappellaro1,Dahlen:2004km,cappellaro2}, or inferred from the
 star formation rate (\sfr), which in turn can be extracted from the
 data on optical or far ultraviolet luminosity of galaxies
 \cite{Cole:2000ea,Glazebrook:2003xu,Baldry:2003xi,Schiminovich:2004km}.
 It can also be constrained from the metal abundances in our local
 universe (see e.g. the discussion in \cite{Strigari:2005hu}).
These different methods have uncertainties of various nature: statistical, systematic, or 
due to theoretical priors.  In particular, the connection between the \sfr\ and the \snr\ relies on a number of 
theoretical inputs, such as the minimum mass required for a star to become a supernova. These inputs are uncertain and thus contribute to the  error on the \snr, and ultimately to the error on the \df.

The second source of error on the \df\ is the uncertainty on the
 neutrino fluxes originally produced inside a supernova.  
To reduce
 this, one can rely on the results of numerical calculations of
 neutrino transport.
An alternative possibility is to
 use the only experimental information available, i.e. the spectra of
 the neutrino events from SN1987A, as was first proposed by
 Fukugita and Kawasaki 
 \cite{Fukugita:2002qw}.

Motivated by theory or by observational results, several authors have
  combined representative neutrino spectra with realistic
 models of the \snr\ to obtain the \df\
 \cite{Totani:1995dw,Malaney:1996ar,Hartmann:1997qe,Kaplinghat:1999xi}.
 Others have investigated the connection with neutrino detection
 \cite{Ando:2002ky,Ando:2002zj,Strigari:2003ig,Ando:2004hc,Ando:2004sb,Cocco:2004ac,Beacom:2005it}
 or the possibility to constrain the \snr\ and the \sfr\ using the
 bound (\ref{sklim}) \cite{Fukugita:2002qw,Strigari:2005hu}.  In most of the  calculations of the
 \df\ available in literature  the
 \snr\ is inferred from the \sfr\ and the neutrino spectra from
 numerical calculations are adopted.  Exceptions are ref. \cite{Fukugita:2002qw}, and ref. \cite{Kaplinghat:1999xi}, 
where the \snr\ was constrained using  the metal enrichment history of the universe.
In all previous works, the  quoted uncertainties on the \df\ are indicative.

In this work I develop the complementary method of Fukugita and Kawasaki.
Specifically, here I analyze the SN1987A data, taking into account neutrino oscillations, to constrain the 
neutrino fluxes in the different flavors produced inside the star.
I also perform a fit of the measurements of the \snr\ from direct \sn\ observations. This is meant to be a first step towards a combined fit of all the available data, and has a value of its own because it is free from the uncertainties that effect the \sfr-\snr\ connection, as emphasized in \cite{Strigari:2005hu}.
Finally, I combine the results of the two analyses, and use them to find the interval of values of the \df\ allowed at a given confidence level.
This study answers the well defined question of what we can conclude on the \df\ if
we decide to rely solely on \emph{direct} experimental information. 
If compared to the previous literature, it shows how the prediction of the \df\ changes with a change of approach in the calculation, and this is important  to provide robust guidance
for experimental searches of this flux.
At a more technical level, this work provides a statistically meaningful error on the \df.
This is relevant 
considering that, as the technology of neutrino telescopes advances,
likely the phase of discovery of a supernova neutrino signal will be
replaced by a phase of detailed analyses and the consideration of
uncertainties on the \df\ predictions will become necessary.

The text is organized as follows: after a section on generalities (Sec. \ref{sec:gen}), I present the analysis of the  SN1987A data in Sec. \ref{87}, and that of the \snr\ measurements in Sec. \ref{snr}.  The combination of the two and the final results for the \df\ are given in Sec. \ref{res}.  Discussion and conclusions follow in Sec. \ref{concl}. 


\section{Generalities}
\label{sec:gen}

Core collapse supernovae are the only site in the universe today 
 where the matter density is large enough  to have the buildup of a
thermal gas of neutrinos.  Thanks to their lack of electromagnetic
interaction, these neutrinos can diffuse out of the star over a time
scale of few seconds, much shorter than the diffusion time of photons. This makes
the neutrinos the principal channel of emission of  the ${\mathcal
O} (10^{53})$ ergs of gravitational energy that is liberated in the
collapse.
 The energy spectrum of each flavor of neutrinos is expected
to be thermal near the surface of decoupling from matter, but then it changes due to propagation effects. 
One of these effects is scattering. 
Numerical modeling indicates that, after scattering right outside the decoupling region, neutrinos  of a given flavor $w$ ($w=e,\mu,\tau$) have energy spectrum \cite{Keil:2002in}:  
\be
 \frac{{d} N_w}{{d} E}\simeq \frac{(1+\alpha_w)^{1+\alpha_w}L_w}
  {\Gamma (1+\alpha_w){E_{0w}}^2}
  \left(\frac{E}{{E_{0w}}}\right)^{\alpha_w}
  e^{-(1+\alpha_w)E/{E_{0w}}},
  \label{nuspec}
\ee
where $E$ is the neutrino energy, $L_w$ is the (time-integrated) luminosity in the species $w$ and $E_{0w}$ is the average energy of the spectrum.  The quantity $\alpha_w$ is a numerical parameter, $\alpha_w \sim 2 - 5$ \cite{Keil:2002in}.  The non-electron neutrino flavors, $\numu$, $\nutau$, $\barnumu$ and $\barnutau$ (each of them denoted as $\nu_x$ from here on), interact with matter more weakly than $\nue$ and $\barnue$, and therefore decouple from matter in a hotter region. This implies that at decoupling  $\nu_x$ has a harder spectrum: $E_{0 x} \gta E_{0 \bar e}$. 
Numerical calculations confirm this, but still leave open the question of how strong the inequality of energies is and of how energetic the neutrino spectra are.  The data from SN1987A are not conclusive on this, as it will appear later. 
Indicative values of the average energies are: $E_{0 \bar e} \sim 12 - 18$ MeV, $E_{0 x}\sim 15 - 22$ MeV. 
Here I consider antineutrinos only, since the $\barnue$  species dominates a detected signal in water, and the contribution of other species is negligible for both the SN1987A data and for a detection of the \df.

The second important effect of propagation on neutrinos is that of
flavor conversion (oscillations). Conversion occurs at matter density of $\sim 10^{3}~{\rm  g\cdot cm^{-3}}$ or smaller (see e.g. \cite{Dighe:1999bi}), where scattering is negligible, due to the interplay of neutrino masses, flavor mixing and coherent interaction of neutrinos with the medium \cite{Mikheev:1986if}
 Thus, the flux of $\barnue$ of energy $E$ in a detector is a linear combination of the original fluxes in the three flavors:
\be
\frac{{d} N^{det}_{\bar e}(E)}{{d} E}=(1+z) \sum_{w=e,\mu,\tau} \frac{{d} N_w(E')}{{d} E'} P_{{\bar  w}{ \bar e}}(E, z)~,
\label{conv}
\ee where I take into account the redshift $z$: here $E' =
E(1+z)$. The factor $P_{{\bar w}{ \bar e}}(E, z)$ is the probability
that an antineutrino produced as  $\bar \nu_w$ is detected as
$\barnue$; it describes the conversion inside the star and
in the Earth and depends on the neutrino mixing matrix and
mass spectrum. In particular, the conversion inside the star depends
on the mass hierarchy (i.e. the sign of the atmospheric mass
splitting, $\Delta m^2_{31}$) and on the mixing angle $\theta_{13}$
(assuming the standard parameterization of the mixing matrix, see
e.g. \cite{Krastev:1988yu}). For inverted mass hierarchy ($\Delta m^2_{31}<0$) the
propagation of antineutrinos is adiabatic
if $\sin^2\theta_{13} \gta 10^{-4}$ \cite{Dighe:1999bi,Lunardini:2003eh}, resulting in a complete
permutation of fluxes: $P_{{\bar \mu}{ \bar e}}+P_{{\bar \tau}{ \bar
e}}=1$. In all the other cases (normal mass hierarchy and/or smaller
$\theta_{13}$) the permutation is only partial. I refer to the
literature for more details \cite{Dighe:1999bi,Lunardini:2003eh,Lunardini:2001pb}.

The contribution of individual supernovae at different redshifts $z$
 to the \df\ at Earth is determined by the cosmic rate of supernovae
 $R_{SN}(z)$, defined as the number of supernovae in the unit of
 (comoving) volume in the unit time.  
The rate at present is $R_{SN}(0)\sim {\mathcal O}(10^{-4})~{\rm Mpc^{-3}~yr^{-1}}$.
Observations as well as theory
 \cite{Hernquist:2002rg} indicate that this rate increases with the
 redshift, meaning that supernovae were more numerous in the past (Sec. \ref{snr}).
 
 By
 combining the flux from an individual supernova with the rate of
 supernovae one finds the flux of $\barnue$ (differential in energy,
 surface and time) in a detector at Earth: 
 \be
\Phi(E)=\frac{c}{H_0}\int_0^{z_{ max}} R_{ SN}(z)\frac{{d}
N^{det}_{\bar e} (E^\prime)}{{d} E^\prime} \frac{{d}
z}{\sqrt{\Omega_{ m}(1+z)^3+\Omega_\Lambda}} 
  \label{flux}
\ee
 (see
 e.g. \cite{Ando:2004hc}). Here $\Omega_{ m}$ and $\Omega_\Lambda$ are the fraction of the cosmic energy density in matter and dark energy respectively;  $c$ is the speed of light and $H_0$ is the Hubble constant.  

The expression (\ref{flux})  is an approximation, because it does not include a number of potentially relevant -- but not well known --  effects.  One of these is the individual variations in the neutrino fluxes emitted, from one star to another,  due to different progenitor mass and type, different amount of rotation and of convection, etc.. 
How large these variations can be is still an open question.  Experimentally, an answer will come by comparing the SN1987A data with those of a future nearby supernova. 
Theoretical studies are not systematic enough to give a correction to Eq. (\ref{flux}).
They hint toward little variation in the spectral shapes of the neutrinos, with possible variations by up to a factor of two in luminosity depending on the mass of the progenitor star  \cite{Thompson:2002mw}.  A second effect not included in Eq. (\ref{flux}) is the deviation from the continuum limit in the supernova rate, due to supernova explosions in a radius of few Megaparsecs from Earth.   These could influence the number of  events recorded in the space  of a few years  \cite{Ando:2005ka}\footnote{I am grateful to A. Gruzinov for directing my attention to this.} 
and constitute an indirect motivation for studying the flux in the continuum limit: indeed, to be able to subtract the contribution of this flux from an observed signal would be important to conclude about a possible supernova event in our galactic neighborhood. 


\section{Neutrino flux from an individual supernova: SN1987A}
\label{87}

As follows from Sec. \ref{sec:gen}, the \df\ in a
detector depends on two sets of parameters. The first refers to the
original spectra of the neutrinos and to conversion effects, while the
second describes the \snr\ function, $R_{SN}(z)$.  In this section I
discuss the constraints on the first set of variables by analyzing the 
SN1987A data. These constraints will be used later in the calculation of
the \df.

\subsection{The data analysis}
\begin{figure}[htbp]
  \centering
    \includegraphics[width=0.75\textwidth]{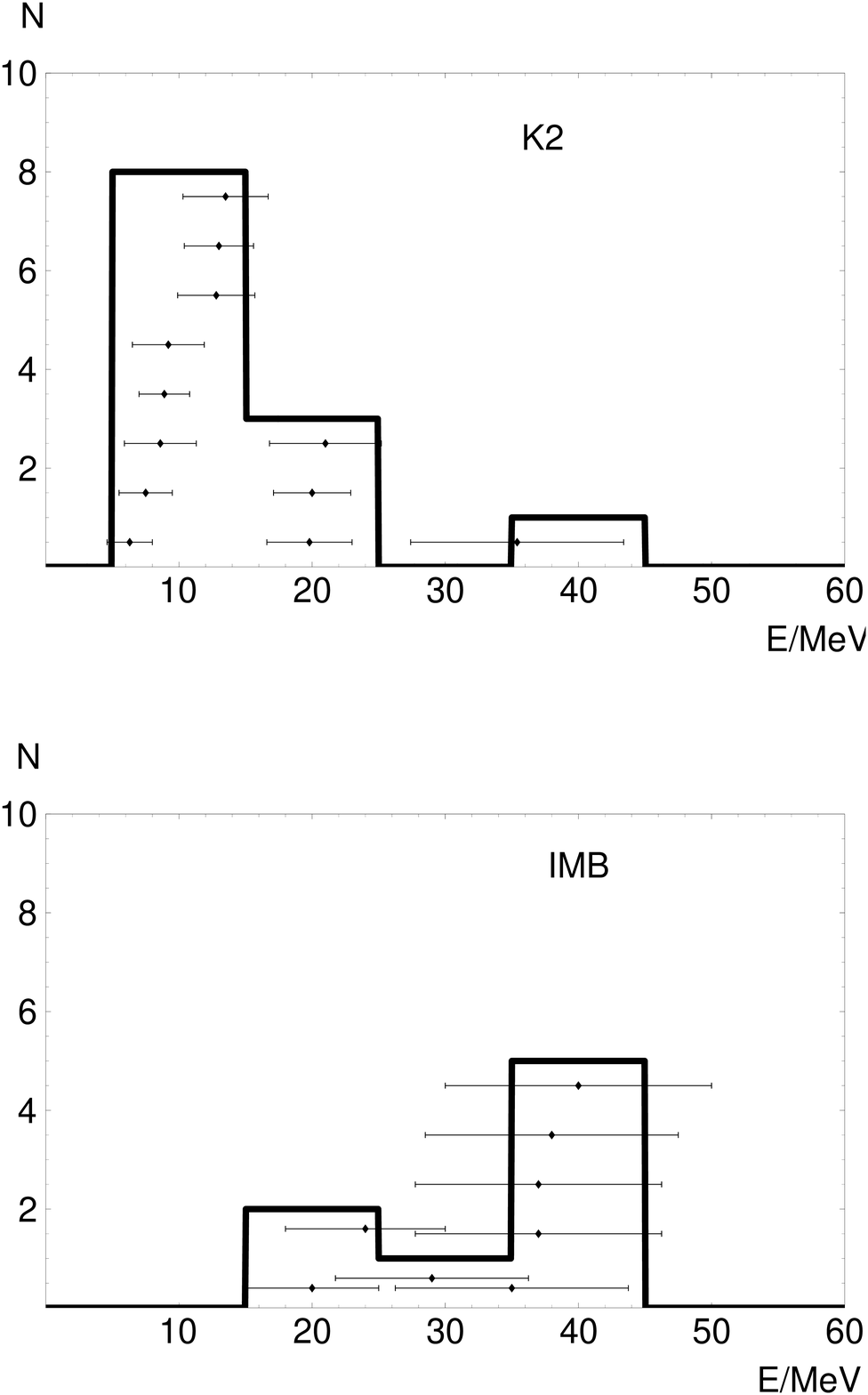}
\caption{The energy spectra of the positrons observed at K2 and IMB. The energies of the individual events with their errors are shown as well, as in ref. \cite{Mirizzi:2005tg}.}
\label{thedata}
\end{figure}

As input, I adopted the twelve data points from Kamiokande-II
\cite{Hirata:1987hu,Hirata:1988ad} and the eight events from IMB
\cite{Bionta:1987qt}, with their errors as published.  I assumed that
all these events are due to the inverse beta decay
$\barnue+p\rightarrow n+e^+$.  
The distributions of the observed positron energies at the two detectors are given in Fig. \ref{thedata}, together with the energies and errors of the individual events.
I took the distance to SN1987A to be
$d_{87}=50 $ kpc, and neglected the $\sim 10\%$ error  on it \cite{mitchelletal}.  The effect of this uncertainty on the \df\ is negligible compared to  the much larger errors of other origin.
 The procedure to calculate the signal at
Kamiokande-II and IMB given a set of parameters follows that of
ref. \cite{Lunardini:2004bj}.  The parameters subject to scan were
five: two luminosities, $L_{\bar e},~L_x$, two average energies
$E_{0 \bar e}$, $E_{0 x}$, and the mixing angle $\theta_{13}$.
I have assumed $\tan^2 \theta_{12}=0.45$, $\Delta m^2_{21}=8 \cdot
10^{-5}~{\rm eV^2}$, as it is given by solar neutrinos and KamLAND
(see e.g. \cite{Aharmim:2005gt,Araki:2004mb}), $|\Delta m^2_{31}|=2.5
\cdot 10^{-3}~{\rm eV^2}$ from atmospheric neutrinos and K2K
\cite{Fukuda:1998mi,Ashie:2004mr,Aliu:2004sq}, and the inverted mass
hierarchy.  The choice to fix the inverted hierarchy and scan over
$\theta_{13}$ includes effectively the case of normal hierarchy, since
for the latter the conversion of antineutrinos is identical to that with inverted hierarchy and complete
adiabaticity breaking in the higher density MSW resonance. I followed refs. 
\cite{Dighe:1999bi,Lunardini:2003eh} for this and other aspects of neutrino conversion in the star and in the Earth.  For definiteness, I fixed
$\alpha_{\bar e}=\alpha_{x}=2.3$, which give a spectral shape close to
Fermi-Dirac, and therefore allow a meaningful comparison with several
other SN1987A data analyses. The final results for the \df\ remain unchanged with the change of $\alpha_{\bar e}$ and $\alpha_{x}$.

I parameterized the density profile of the
progenitor star (necessary to calculate the matter-driven flavor conversion) as $\rho(r)= 4 \cdot 10^{13} (10~{\rm Km}/r)^{3}~{\rm g\cdot cm^{-3}}$,
with $r$ being the radial distance from the center.
For simplicity, I ignored the uncertainty
on $\rho(r)$, as well as the errors on the values of the solar
and atmospheric oscillation parameters. The inclusion of these
uncertainties has no appreciable effect on the final results for the
\df.  I have not included late time shockwave effects \cite{Schirato:2002tg}, as their impact on a time integrated signal is negligible with respect to the large statistical errors and other uncertainties (see e.g. \cite{Ando:2002zj} for this particular aspect).  
The experimental parameters, 
such as efficiency curves and energy resolution functions, are
as in \cite{Lunardini:2004bj}, and the detection cross section was taken from \cite{Strumia:2003zx} (Eq. (25) there).

Given the sparseness of the data, the maximum likelihood method of
analysis is the most appropriate.   
Following   Jegerlehner, Neubig and Raffelt \cite{Jegerlehner:1996kx}, I obtain the likelihood function,
${\mathcal L}_{87}(E_{0 \bar e},E_{0 x},L_{\bar e},L_x,$ $\sin^2 \theta_{13})$, and the quantity $\chi^2_{87}$:
\be
\chi^2_{87}\equiv -2 \ln {\mathcal L}_{87}~.
\label{chi2}
\ee
 Once
the  minimum of $\chi^2_{87}$, $\chi^2_{87,min}$, has been found, the
five-dimensional region of parameters that are allowed at a given confidence level
(C.L.) is given by:
\be
\Delta \chi^2_{87} \equiv \chi^2_{87}(E_{0 \bar e},E_{0 x},L_{\bar e},L_x,\sin^2 \theta_{13}) - \chi^2_{87,min}\leq \chi_5~,
\label{lik}
\ee where $\chi_5=5.86, 9.24, 15.09$ for $68,90,99\%$ C.L..  Like in
\cite{Jegerlehner:1996kx}, here the value of $\chi^2_{87}$ will be
given up to a constant, which is irrelevant for parameter estimation,
according to Eq. (\ref{lik}).  I emphasize that the maximum
likelihood method does not involve any binning of the data \cite{Jegerlehner:1996kx}.  Thus, the
bins used in figures \ref{thedata} and \ref{histo} are only for
illustration and do not influence the likelihood function.

The scan was performed in the
five-dimensional box given by $E_{0 \bar e}=3 - 30$ MeV, $E_{0 x}=3 - 30$ MeV,
$L_{\bar e}=(1.5 - 45)\cdot 10^{52}$ ergs, $L_{x}=(1.5 -
45)\cdot 10^{52}$ ergs, $\sin^2\theta_{13}=10^{-7} - 10^{-2}$. The latter interval covers all the possibilities of spectrum permutation due to conversion in the star \cite{Lunardini:2003eh}.
\emph{No hierarchy of neutrino spectra} was imposed a priori. Results
with certain  priors  will be discussed briefly
later.  

The five-parameters analysis adopted here is statistically meaningful
given the number (20 in total) of data points available.  It is
original, because it combines generality (the absence of theoretical
priors) and great detail in the inclusion of neutrino conversion
effects \footnote{The statistical analyses in the past literature
(e.g. \cite{Jegerlehner:1996kx,Lunardini:2000sw,Minakata:2000rx,Kachelriess:2000fe,Barger:2002px,Mirizzi:2005tg}) used
priors. Most of them had only two fit parameters, with fixed flavor
conversion pattern and fixed ratios of average energy and of
luminosities in $\barnue$ and $\nux$.  }. It is known that a smaller
number of parameters is enough to well reproduce the SN1987A data
\cite{Mirizzi:2005tg} and thus would probably suffice to predict the
\df.  Still, the five-parameters method has been preferred here
because, aside from the specific connection to the \df, it serves a
more general purpose: to give a state-of-art answer to the question of
what we know on the neutrino energies and luminosities \emph{at the
production site}. This question is important in connection with other
physics inside the supernova (neutrino transport, R-process
nucleosynthesis, etc.).  Another reason for the choice is that this
method includes the case when the observed neutrino spectrum is
strongly distorted with respect to an effective thermal
distribution. This can happen if the original $\barnue$ and $\nux$
spectra have very different average energies and luminosities (see
Sec. \ref{separate} and Fig. \ref{histo}).

\subsection{Analyzing Kamiokande and IMB individually} 
\label{separate}

\begin{figure}[htbp]
  \centering
    \includegraphics[width=0.75\textwidth]{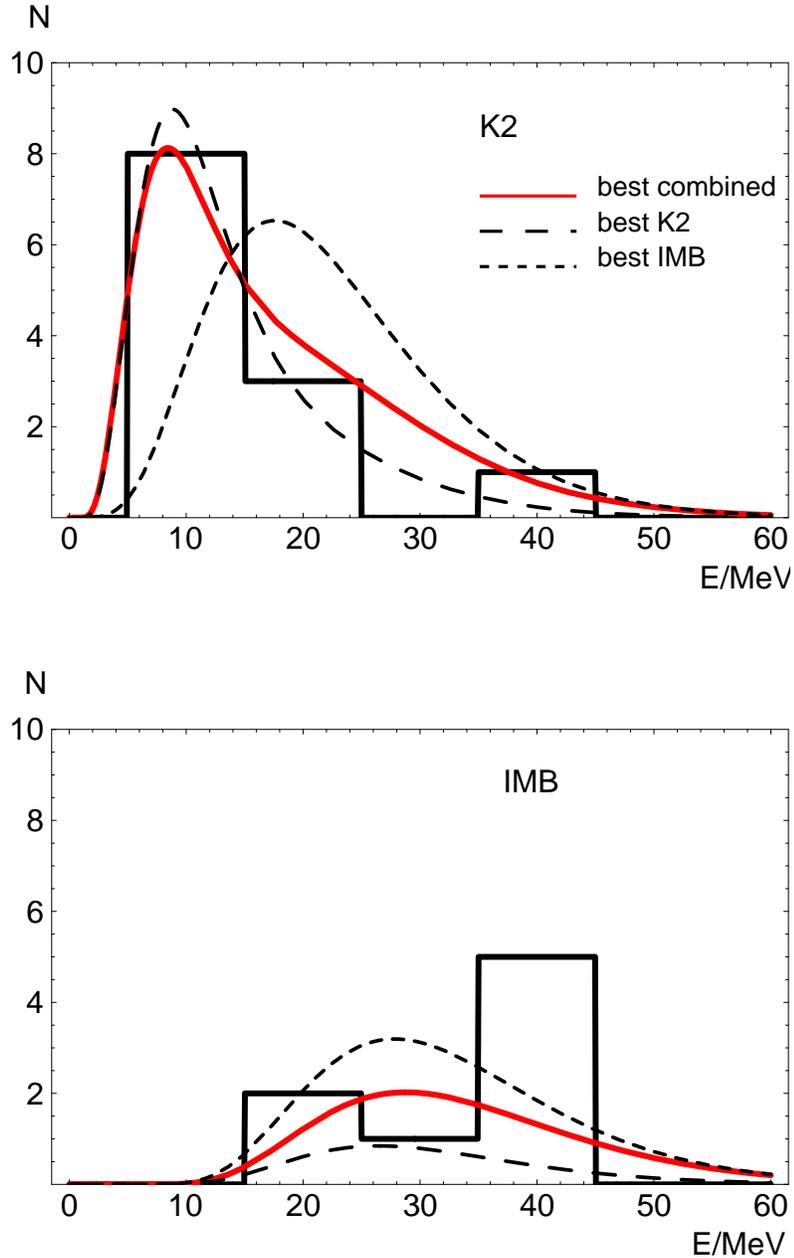}
\caption{The observed  energy spectra of events at K2 and IMB, as in Fig. \ref{thedata}, compared with the predicted spectra in the points of minimum $\chi^2$ for K2 only, IMB only and combined K2 and IMB data sets. The values of the parameters in these points are given in Table \ref{table87}.} 
\label{histo}
\end{figure}

It is useful to consider the K2 and IMB data sets separately first.
Their energy spectra are shown in Fig. \ref{histo}, and their average
energies are summarized in Table \ref{table87}.  One can see that the
two spectra differ substantially.  The differences are in good part
due to different experimental settings (higher energy threshold for
IMB, different volumes, energy resolutions and detection
efficiencies).  Still, on top of these technical differences, a
tension exists between the two spectra \footnote{The angular
distribution of the events at both detectors is only marginally
compatible with the predictions for inverse beta decay data, suggesting the presence of an anomaly in the
data \cite{LoSecco:1988hb}.  The standard interpretation of
the SN1987A signal in terms of inverse beta dacay is not excluded,
however, and therefore it is adopted here.}. Specifically, the fact
that the IMB spectrum has a maximum at about 40 MeV contrasts with the
K2 signal, which is compatible with an exponentially decreasing
spectrum in that energy region.

 In the light of this tension,  it is meaningful to study the $\chi^2$ functions of K2, $\chi^2_{K2}$,  and IMB, $\chi^2_{IMB}$,  separately to understand how each data set influences the combined function $\chi^2_{87}$. 
 
 As expected given the large number of fit parameters, both $\chi^2_{K2}$ and $\chi^2_{IMB}$ have many degenerate minima, that correspond to the same spectrum of events.   Two of these minima are given in Table \ref{table87} for illustration.   From these sets of minima, one can see that:

 \begin{itemize}
 
 \item The K2 observed spectrum is best reproduced by a superposition of two original spectra, one very soft, $E_{0}\sim 4 -5 $ MeV and the other more energetic: $E_{0}\sim 12 -14 $ MeV.   The soft component has to have very high luminosity, $L \sim 3 \cdot 10^{53}$ ergs,   higher by a factor of several with respect to the hard one.   In this composite spectrum, the soft part accounts for the peak of the K2 data in the lowest energy bin, $E \sim 10$ MeV, while the harder part reproduces the tail of the observed spectrum at higher energy.  This is illustrated in Fig. \ref{histo}.  The scenario favored by the K2 data  can be realized for normal hierarchy, or inverted hierarchy with small $\theta_{13}$, a soft $\barnue$ original spectrum and a harder $\nux$ one (the case with soft $\nux$ and hard $\barnue$ fits equally well, but is not theoretically motivated).  
One must be aware, however, that this spectrum contrasts with the theory (Sec. \ref{sec:gen}) because  the soft component has too low energy and too high luminosity, comparable to the total energy output predicted for a \sn\
\footnote{The conclusion of ref. \cite{Lunardini:2004bj}, that the K2 data alone do not favor a composite spectrum, is not in conflict with the results of this work.  It reflects the more conservative assumptions adopted in \cite{Lunardini:2004bj} about the original neutrino fluxes, such as the condition of comparable luminosities: $L_x/L_{\bar e } = 0.5 - 2$.}.

 \item  The IMB data favor a purely thermal spectrum --  instead than  a superposition of thermal spectra -- with average energy $E_0 \sim 13 - 15$ MeV and luminosity $L \sim 0.5 \cdot 10^{53}$ ergs,  in acceptable agreement with theory.  These parameters are determined by the average energy of the data and by the number of events.   A thermal spectrum is favored over a composite one because of its smaller width  (see e.g. \cite{Lunardini:2004bj} for a detailed discussion), that better reproduces the narrow spectrum of the IMB data.   The spectral characteristics favored by IMB are realized if the mass hierarchy is inverted with adiabatic conversion (large $\theta_{13}$), for which the whole $\barnue$ signal is due to the original $\nux$ flux.    The other cases, where the conversion is only partial,  fit equally well  if the original $\nux$ and $\barnue$ fluxes have the same spectrum or if one of them is suppressed with respect to the other one. The suppression can be due  either to small luminosity or  to average energy much below the IMB detection threshold ($E_{th}\sim 20$ MeV \cite{Bionta:1987qt}). The latter point is especially important, as will be seen. 
 
 \end{itemize}

 \begin{table*}
\centering
\begin{tabular}{| l | l | l | l | l | l |}
\hline
\hline
&  data & best K2 & best IMB & best combined & 68\% C.L. combined \\
\hline
\hline 
 $\sin^2\theta_{13}$ & &  $10^{ -7}$  & $10^{-2}$ & $10^{ -7}$  & $10^{ -7}-10^{ -2}$ \\
\hline
 $E_{0 \bar e}/{\rm MeV}$ & &  4.6 &  unconstrained & 4.2 & see Fig. \ref{1987aplot} \\
\hline
 $E_{0x}/{\rm MeV}$ & & 12.7 & 13.6 & 14.9  & see Fig. \ref{1987aplot} \\
\hline
 $L_{e}/{\rm 10^{53}~ergs}$ & &  3.4 & unconstrained & 4.4 & unconstrained \\
\hline
 $L_{x}/{\rm 10^{53}~ergs}$ & & 0.51 & 0.45 & 0.8 & see Fig. \ref{1987aplot} \\
\hline
 $\chi^2_{K2}$ & & 42.1 & 54.7& 43.8 & \\
\hline
 $\chi^2_{IMB}$ & & 49.5 & 39.1 & 40.4 & \\
\hline
 $\chi^2_{87}$ & & 91.6 & 93.8 & 84.2 & \\
\hline
 $N_{K2}$ & $12.0 \pm 3.5 $ & 12.1 &  14.3 & 14.6 & 7.8 - 24.2 \\
\hline
 $N_{IMB}$  &  $8.0 \pm 2.8 $ & 2.0  &  8.1 & 5.4 & 2.4 - 10.3 \\
\hline
 $\langle E \rangle_{K2}/{\rm MeV}$ & $14.7 \pm 1.1$ & 13.2 & 29.7 & 16.7 & 13.5 - 21.2 \\
\hline
 $\langle E \rangle_{IMB}/{\rm MeV}$ & $31.9 \pm 2.3 $ & 29.7 & 31.2 & 32.6 & 28.4 -  38.3 \\
\hline
\hline
\end{tabular}
\caption{Summary of relevant quantities in the points of maximum likelihood of K2 and IMB individually, as well as in the point of maximum combined likelihood and in the 68\% C.L. allowed region.  The three specific points used here were chosen for illustration among the degenerate maxima of the likelihood function (minima of $\chi^2$). The values of $\chi^2$ are given up to a constant (irrelevant for parameter estimation).
Where meaningful, the values given by the data are presented too (from ref. \cite{Lunardini:2004bj}).}
\label{table87}
\end{table*}

\subsection{Combined analysis: results}

 Comparing the neutrino spectra favored by K2 and by IMB separately,
 one infers that a good combined fit exists.  The key to see this is
 to observe that the average energy favored by
 IMB is similar to that of the hard component
 of the spectrum favored by K2 (see Table \ref{table87}).  Moreover, a
 very soft spectral component would affect IMB only marginally, due to
 threshold effects, as was mentioned in Sec. \ref{separate}.  Thus,
 one expects that the combination of the two data sets will favor a
 composite soft+hard spectrum qualitatively similar to that favored by
 the K2 data.  The closeness to the K2-only result is motivated also
 by the fact that the K2 data dominate the statistics.

Some features of the $\chi^2_{87}$ function can be predicted as well. One
expects degeneracies, even though not as extended as in the case of K2
and IMB separately.  In particular, a degeneracy in  $\chi^2_{87}$
should exists between the region with $E_{0\bar e } < E_{0x}$ and that
with $E_{0\bar e } > E_{0x}$. The reason lies in the fact that, if the
neutrino conversion is only partial (probability $0<P_{\bar x  \bar e}<1$), the neutrino
spectrum at Earth is symmetric under the transformation 
$ P_{{\bar  e}{ \bar e}} {d} N_{\bar e}(E')/{d} E' \leftrightarrow (1-P_{{\bar  e}{ \bar e}}) {d} N_{\bar x}(E')/{d} E' $
(see Eq. (\ref{conv})).  Such symmetry
implies an approximate symmetry (broken by effects of oscillations in
the Earth and by the energy dependence of $P_{{\bar  e}{ \bar e}}$) between the $E_{0\bar
e } < E_{0x}$ and $E_{0\bar e } < E_{0x}$ portions of the parameter
space.

\begin{figure}[htbp]
  \centering
 \includegraphics[width=0.990\textwidth]{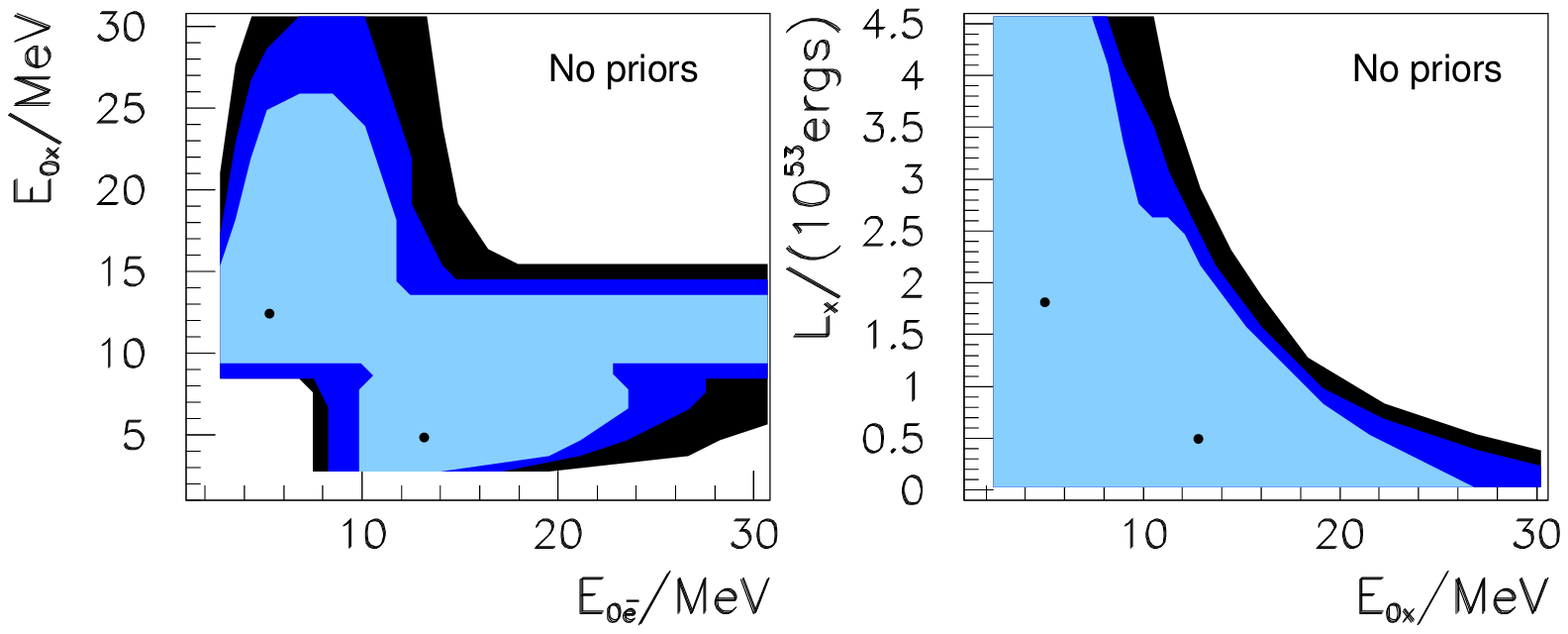}
\includegraphics[width=0.990\textwidth]{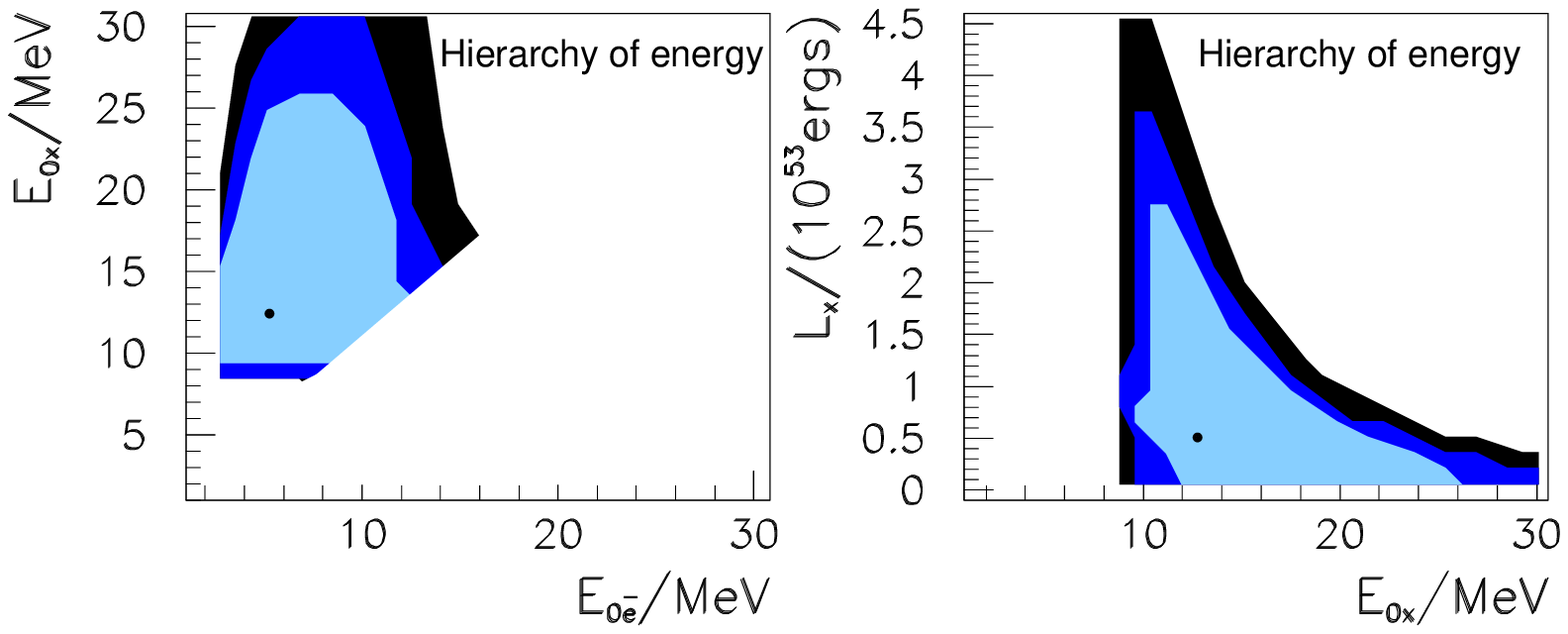}
\caption{Projections of the 68\%,90\%,99\% C.L. regions allowed by the SN1987A data on the planes  $E_{0 \bar e}-E_{0x}$ and   $E_{0 x}-L_{x}$, without any prior (upper panels) and with the hierarchy $E_{0 \bar e}<E_{0 x}$ (lower panels). The dots in each panel mark the projections of the points of maximum likelihood (see Table \ref{table87}).  The entire plane  $E_{0 \bar e}-L_{\bar e}$ (not shown) is allowed at 68\% C.L..   }
\label{1987aplot}
\end{figure}

The likelihood analysis, summarized in Table \ref{table87} and Figs. \ref{histo} and \ref{1987aplot}, confirms the intuitions.  The predicted degeneracy is observed.
The $\chi^2_{87}$ has two minima in the points
$(\log(\sin^2\theta_{13}),E_{0 \bar e},E_{0 x}, L_{\bar e}, L_{x})=(-7,
4.6~{\rm MeV},12.7~{\rm MeV},3.4 \cdot 10^{53}~{\rm ergs},0.51 \cdot
10^{53}~{\rm ergs})$ ($\chi^2_{87}=84.1$) and
$(\log(\sin^2\theta_{13}),E_{0 \bar e},E_{0 x}, L_{\bar e},
L_{x})=(-3.6, 12.5~{\rm MeV},5.2~{\rm MeV},6 \cdot 10^{53}~{\rm
ergs},1.7 \cdot 10^{53}~{\rm ergs})$ ($\chi^2_{87}= 84.7$).  They
correspond to the type of spectrum preferred by K2.  Here, the same
considerations done for the K2-only fit apply regards the conflict
between the best fit scenario and theory.  It is important to stress,
however, that the $\chi^2_{87}$ function is very shallow, with many
points having essentially the same goodness of fit as the minimum.
The allowed region in the parameter space is very extended and
includes theoretically motivated scenarios at 68\%
C.L. (Fig. \ref{1987aplot}).  The region is compatible with the
constraints on the neutrino spectrum put by light element synthesis
during galactic chemical evolution \cite{Yoshida:2005uy,Yoshida:2006qz} (see also the related works \cite{Woosley:1988ip,Haxton:1997gk}), which
favor rather soft muon and tau neutrino spectra: $E_{0x} \sim 15 - 18$ MeV
\cite{Yoshida:2005uy}.  A detailed comparison of those constraints with the results
of this paper is not possible, however, due to the different working assumptions used here.

\begin{figure}[htbp]
  \centering
 \includegraphics[width=0.6\textwidth]{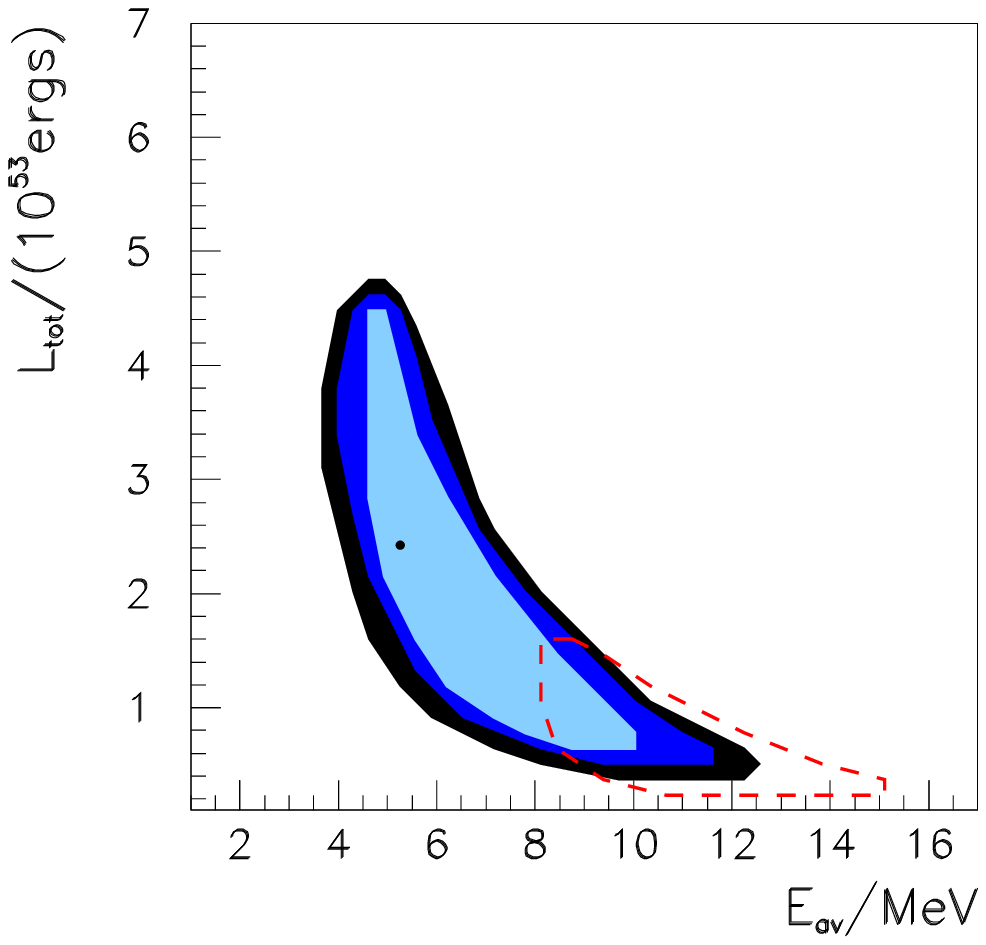}
\caption{ The best fit point and the regions allowed by the SN1987A data at 68\%,90\%,99\%
C.L. in the  space of the average energy and luminosity
of the $\barnue$ flux at Earth (after conversion in the star). For comparison, I show the 99\% C.L. region found in ref. \cite{Mirizzi:2005tg} (dashed contour). There the same flux was described by a spectrum of the form (\ref{nuspec}). } 
\label{comparemirizzi}
\end{figure}

The upper panels of Fig. \ref{1987aplot} illustrate the details of the allowed region.   There I give the
projections on the $E_{0 \bar e}-E_{0 x}$
and $E_{0 x}-L_x$ planes of the 5D allowed regions given by
Eq. (\ref{lik}) for different confidence levels, together with the
projections of the points of maximum likelihood. 
The figure confirms that a rather large portion of the parameter space
is allowed, with better sensitivity to the $\nux$ flux with respect to
the $\barnue$ one.  Unless assumptions are made on the oscillation
pattern, the $\barnue$ flux is unconstrained, due to the possibility
that the mass hierarchy be inverted with perfectly adiabatic
conversion ($\sin^2 \theta \gta few \cdot 10^{-4}$) inside the
star. Under such conditions the original $\barnue$ flux is completely
converted into $\nux$ and does not affect the observed $\barnue$
signal.  The $\nux$ flux is constrained loosely: every value of
$E_{0x}$ gives a good fit provided that the other parameters are
suitably adjusted, and a similar statement is valid for $L_x$.  Notice
that the $\nux$ luminosity and average energy are not constrained from
below: indeed, they are allowed to be zero, in the case when the
$\barnue$ spectrum alone gives a good fit to the data.  This can
happen for all conversion patterns except the one with complete
$\barnue - \nux$ permutation, and requires $E_{0\bar e}\sim 13 -15
$ MeV.  In all cases, the region with $E_{0\bar e } \lta
5$ MeV and $E_{0x}\lta 5$ MeV is excluded because it gives a very soft
neutrino spectrum, with a too small (or even vanishing) number of
events at IMB.

If the
analysis is restricted to the region motivated by the theory, $E_{0\bar e } < E_{0x}$ (lower panels of Fig. \ref{1987aplot}),
some conclusions change:
$E_{0x}$ must be larger than $\sim 10$ MeV and $E_{0\bar e}$ can not exceed $\sim 16$ MeV. 

The Table \ref{table87} gives further details on one of the best fit points, confirming that it reproduces the data well in the number of events, $N_{K2}$ and $N_{IMB}$ and average energies, $\langle E \rangle_{K2}$ and $\langle E \rangle_{IMB}$.
The goodness of the fit also appears from the superposition of the best fit spectrum with the data, in Fig. \ref{histo}.

It is informative to present the results in a complementary way:
by using a smaller number of variables and choosing parameters that
describe the $\barnue$ flux at Earth, instead than at the production
point.  Here I choose the average energy and luminosity of such flux,
$E_{av}$ and $L_{tot}$. Fig. \ref{comparemirizzi} shows the allowed
region in the space of $E_{av}$ and $L_{tot}$, obtained after
marginalizing the likelihood function over the three parameters
orthogonal to these. For comparison, in the figure I also show an
example of result obtained by fitting the data with two parameters
only (i.e., keeping the other three parameters fixed instead than
marginalizing over them). The example is taken from
ref. \cite{Mirizzi:2005tg} (fig. 2 there), where the $\barnue$ flux at
Earth is described by a spectrum of the form (\ref{nuspec}).  The
marginalized region confirms the finding of a particularly soft and
luminous $\barnue$ flux being favored by the data.  The region is
compatible with but more extended than the two-parameters result from
ref. \cite{Mirizzi:2005tg}. This reflects the fact that a
five-variables parameterization includes neutrino spectra that fit the
data well but are not reproduced with a smaller number of
parameters. It is confirmed, therefore, that five-dimemsional and the
two-dimensional methods are not equivalent. 


\section{The supernova rate}  
\label{snr}

The cosmic rate of supernovae,
 $R_{SN}(z)$,   can be inferred from astrophysical data in a variety of ways.
  Here I review two:

\begin{enumerate}

\item  \underline{Measurements of the \snr\  } from observations of core collapse \sne. This is the most direct method.  To date, four measurements of the \snr\ have been made in this way, covering the interval of redshift $z \sim 0 - 0.9$ \cite{cappellaro1,Dahlen:2004km,cappellaro2}. They are summarized in Table \ref{snrtable} and in Fig. \ref{snrband}.  While the connection between the observations and the \snr\ is immediate, in principle, one should keep in mind that the results are affected by dust obscuration (extinction) and possible misidentification of the observed objects.   
These effects can be modeled theoretically and subtracted.
Three of the measurements in Table \ref{snrtable} have been corrected in this way.   The point at $z=0.26$, indicated with an empty circle in Fig. \ref{snrband}, is not corrected for extinction and for this reason it will not be included in the analysis here. 

\item \underline{measurements of the star formation rate (\sfr)}, $R_{SF}(z)$.  From these, the \snr\ is found through the equation:
\begin{eqnarray}
 R_{ SN}(z)&\simeq &\frac{\int_{8 M_\odot}^{50 M_\odot}d m~\phi(m)}
  {\int_{0}^{125 M_\odot}d m~m\phi(m)}R_{SF}(z)~\sim 10^{-2} M^{-1}_\odot R_{SF}(z)~,
  \label{eq:SNrate}
\end{eqnarray}
(see e.g. \cite{Ando:2004hc})
 where $\phi(m)$ is the initial mass function, decreasing roughly as a power  $-2$ of the mass $m$, and $M_\odot$ is the mass of the Sun, $M_\odot\simeq 1.99 \cdot 10^{30}$ Kg.
  The limits of integration represent the interval of mass for which a star becomes a core collapse supernova.  
 With respect to direct searches of \sne, this method has the advantage of higher statistics, since several measurements of the \sfr\ are available, extending up to $z\sim 6-7$ (see e.g. \cite{Cole:2000ea,Glazebrook:2003xu,Baldry:2003xi} and \cite{Hopkins:2006bw} for an overview and further references).   On the other hand, the \snr\ obtained with this approach is indirect: it is affected by a number of theoretical uncertainties through the relation (\ref{eq:SNrate}), such as those on the initial mass function and on the interval of progenitor masses.   The lower mass cut in (\ref{eq:SNrate}) influences the normalization of the \snr\  strongly.  I refer to \cite{Hopkins:2006bw} for an in-depth discussion of these aspects.   On top of the uncertainties mentioned, the \sfr\ measurements are  affected by extinction, like the direct \sn\ observations.  Likely, the largest theoretical error is associated to the normalization of the \snr, however the possibility that the ratio of \sfr\ and \snr\ be redshift-dependent -- thus violating Eq. (\ref{eq:SNrate})--  is not excluded. This is a possible source of error as well.
  
\end{enumerate}

Clearly, the two methods are complementary.  Their results agree in the basic facts: the \snr\ today is of order $R_{SN}(0)\sim 10^{-4} ~{\rm yr^{-1} Mpc^{-3}}$.  The rate
increases with $z$ and is consistent with a broken power law:
\begin{eqnarray}
{ R}_{{ SN}}  (z) & =  & { R}_{{ SN}}(0) (1+z)^\beta \quad   
  {\rm for} \quad  z <
       1  \nonumber \\  & = & { R}_{SN}(0) 2^{\beta -\alpha} \,  
(1+z)^\alpha  \quad  {\rm for}
       \quad z > 1~  .
\label{eq:powerlaw}
\end{eqnarray}
The values of ${ R}_{{ SN}}(0)$ favoured by the two approaches are somehow different, with method (2) giving a rate larger by a factor of $\sim 2 -3 $ with respect to method (1). This is probably a manifestation  of the uncertainties that affect both methods.  For example, it was checked that the discrepancy disappears if the lower mass cut in (\ref{eq:SNrate}) is increased to $10M_\odot$ \cite{Hopkins:2006bw}. 

In this work I use method (1), motivated by its being direct and therefore more robust, and suitable for a conservative estimate of the \df.   For this  specific application,
its being limited to small resdhift is of little consequence, since
the \df\ above realistic energy thresholds is dominated by \sne\
closer than $z \sim 1$ \cite{Ando:2004hc}.   For this reason, here I fix $\alpha=0$, a value favored by the \sfr\ data  \cite{Hopkins:2006bw}.
 I perform a maximum likelihood analysis of the three extinction-corrected measurements of the \snr\ given in Table \ref{snrtable}, to find  the
allowed region of the parameters $R_{SN}(0)$ and $\beta$  of Eq.  ({\ref{eq:powerlaw}}).  
In the approximation of gaussian errors, I calculate the likelihood function 
 ${\mathcal L}_{SNR} (R_{SN}(0), \beta) $ and  $\chi^2_{SNR} \equiv -2  \ln {\mathcal L_{SNR}}$. 

\begin{table*}
\centering
\begin{tabular}{| l | l | l | }
\hline
\hline
 reference & redshift $z$ & $R_{SN}/(10^{-4}~{\rm yr^{-1} Mpc^{-3}})$ \\
\hline
\hline 
\cite{cappellaro1}  &  0  & $0.59 \pm 0.24$ \\
\hline
 \cite{cappellaro2}  &  0.26  &  $1.82^{+0.69}_{- 0.56}$ \\
\hline
\cite{Dahlen:2004km}  &  0.3 (average)  & $2.51^{+0.88}_{- 0.75}$ \\
\hline
\cite{Dahlen:2004km}  &  0.7 (average)  & $3.96^{+ 1.03}_{- 1.06}$  \\
\hline
\hline
\end{tabular}
\caption{The measured supernova rate at different redshifts.  The errors represent the $68\%$ confidence level intervals. The values of the redshift in the third and fourth row are averages over bins of redshift.  The point at $z=0.26$ is not corrected for exctinction, while the other three points are.}
\label{snrtable}
\end{table*}

\begin{figure}[htbp]
  \centering
\includegraphics[width=0.900\textwidth]{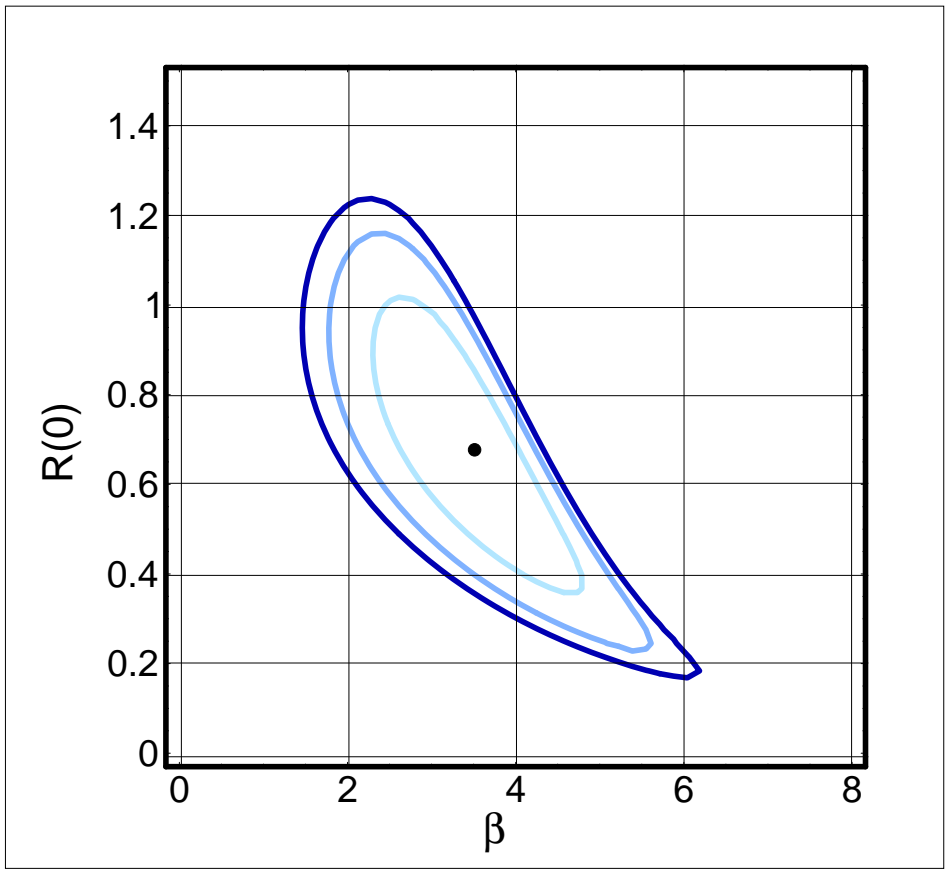}
\caption{Best fit point and isocontours of $\chi^2$ in the space of the parameters describing the \snr\ function, $R_{SN}(z)$. These are the intercept, $R_{SN}(0)$ (in units of $10^{-4}~{\rm yr^{-1}~Mpc^{-3}}$) and the power, $\beta$.  The contours refer to 68.3, 90, 95.4\% C.L.. }
\label{snrcontours}
\end{figure}

\begin{figure}[htbp]
  \centering
\includegraphics[width=1.2\textwidth]{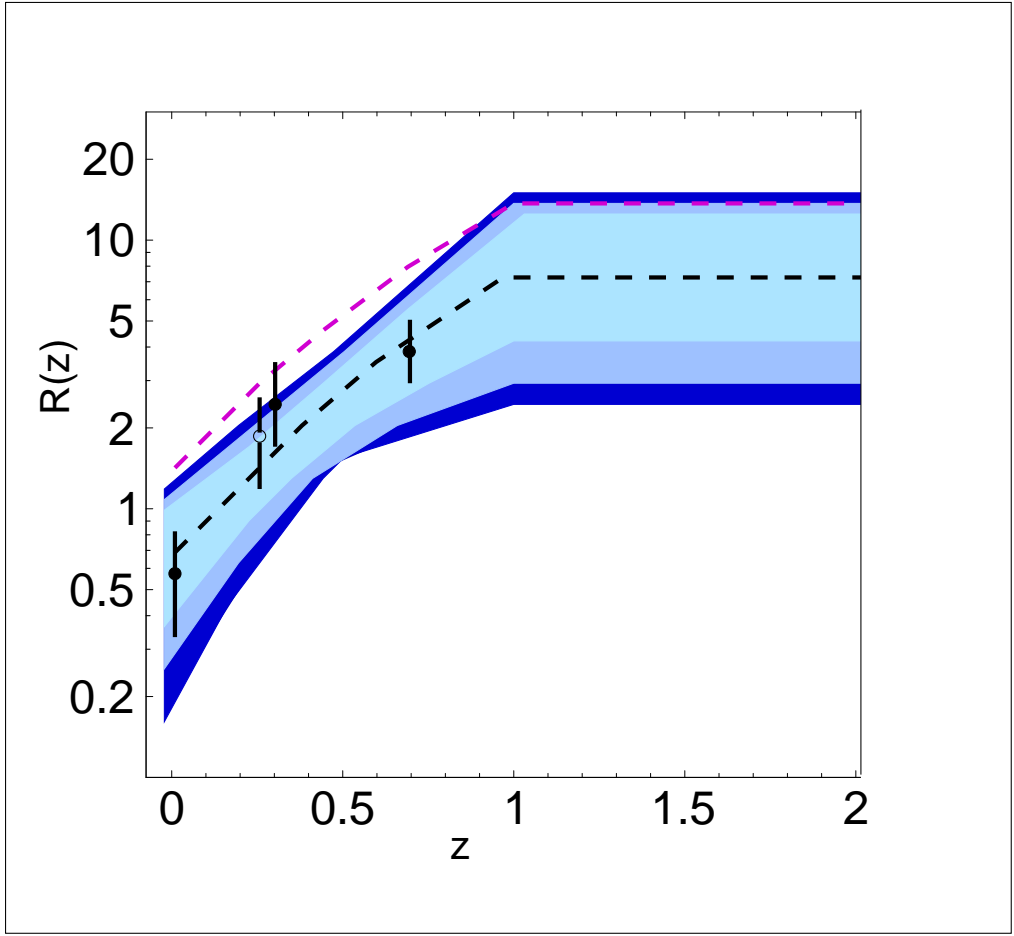}
\caption{The \snr\  (in units of $10^{-4}~{\rm yr^{-1}~Mpc^{-3}}$) as a function of the redshift, $z$.  The colored (shaded) bands are the families of curves having parameters within the 68.3, 90, 95.4\% C.L. contours of Fig. \ref{snrcontours}. The lower dashed line refers to the best fit parameters of my analysis. The measurements of the \snr, with $1~\sigma$ error bar, are shown (from Table \ref{snrtable}). Of these, the point marked with an empty circle was not included in the analysis because it is not corrected for exctinction. The upper dashed line results from using, instead than core collapse data, the \sfr\ data with a specific initial mass function (from ref. \cite{Hopkins:2006bw}, see text).
}
\label{snrband}
\end{figure}

The results are shown in figs. \ref{snrcontours} and \ref{snrband}.
The likelihood is maximal in the point $R_{SN}(0)=0.67 \cdot 10^{-4}~{\rm yr^{-1} Mpc^{-3}}$, $\beta=3.44$, where the value of $\chi^2_{SNR}$ is $\chi^2_{SNR,min}= 3.59$.
Fig. \ref{snrcontours}  shows the point of maximum likelihood and the contours  defined by $\chi^2_{SNR} - \chi^2_{SNR,min}=2.3,4.61,6.17$, corresponding to 68.3, 90, 95.4\% confidence level (C.L.) \footnote{One should keep in mind that these confidence levels are not exact, as they apply rigorously only to  likelihood functions that are perfectly  gaussian in the parameters of interest.}. 
The results of the likelihood analysis confirm that the fit of the data with a power law curve is satisfactory. The allowed region is rather extended, with  $R_{SN}(0)$ and $\beta$ varying by a factor of six and three respectively. 
In Fig.  \ref{snrband} the \snr\ functions allowed at given confidence
level (shaded bands) and the best fit function are compared with the
input measurements of Table \ref{snrtable}.   The figure also gives
the \snr\ function found in ref. \cite{Hopkins:2006bw} using method (2), with the
modified Salpeter B initial mass function (Table 2  in  \cite{Hopkins:2006bw}) and the interval of
progenitor masses as in Eq. (\ref{eq:SNrate}).  Confirming what
already mentioned above, this curve has very similar logarithmic slope,
$\beta=3.35$, but normalization higher by a factor of $\sim 2$ with
respect to the best fit function found in this work.

In connection to this, one may consider performing a global analysis
of the direct \snr\ measurements and of the \sfr\ data together.  Such
analysis warrants a separate work, necessary to make the different
data sets compatible with each other; this in turn requires, among
other things, a careful study of the connection between \snr\ and
\sfr, and an evaluation of the different methods used by different
authors in their statistical analyses and in the correction for dust
obscuration.  This has not been done so far and represents the next
step with respect to this paper. In consideration of the uncertainties
on the normalization of the \sfr, I expect that the combination with
the \sfr\ data would improve the constraint on $\beta$, with only little improvement on $R_{SN}(0)$.

Studies have indicated  that the $z=0$ measurement
used here may underestimate the local \snr\ by a factor of two or so \cite{Mannucci:2003st,Ando:2005ka}.  In
absence of a new determination of the local rate, one can only give
examples of how the likelihood results would change if the $z=0$ point
was higher.  If I rescale this point and its error by a factor of two
(three) the new point of maximum likelihood is $R_{SN}(0)=1.26 \cdot
10^{-4}~{\rm yr^{-1} Mpc^{-3}}$, $\beta=2.21$ ($R_{SN}(0)=1.72 \cdot 10^{-4}~{\rm yr^{-1} Mpc^{-3}}$,
$\beta=1.55$), with only minor change in the value of $\chi^2_{SNR, min}$.  As will be shown, the
effect of a higher $R_{SN}(0)$ is partially compensated by that of a
smaller $\beta$, so that the impact of this change on the \df\ is only
of tens of per cent.


\section{Diffuse neutrino flux}
\label{res}

\subsection{The calculation}

In the framework adopted here, the \df\ depends on seven parameters: five describing the neutrino flux (two average energies, two luminosities and one mixing angle) and two describing the \snr\ (the rate at $z=0$ and the power, $\beta$).  To  calculate the diffuse flux and the uncertainty on it, it is necessary to combine  the information from both the SN1987A and the \snr\  data sets consistently. 

I perform this combination as follows: 

\begin{itemize}

\item I obtain the total likelihood of the neutrino and \snr\ data together, ${\mathcal L}_{tot}$.  Since the two sets of data are uncorrelated,  this is done by simply multiplying the two individual likelihoods. The combined $\chi^2$ is the sum of the two pieces: $\chi^2 = \chi^2_{87}+\chi^2_{SNR}$. 

\item Using Eq. (\ref{flux})  I compute the likelihood for the $\barnue$ diffuse flux in a detector, $\Phi$, by marginalizing ${\mathcal L}_{tot}$ with respect to all the six quantities orthogonal to $\Phi$.  
If the likelihood function  is close to a gaussian near its maximum,  for a given $\Phi$ the  marginalized $\chi^2$ --  called $\chi^2_\Phi$ from here on -- is well approximated by  the global $\chi^2$ minimized with respect to the other six  variables. 
Here I adopt this approximation, widely used in data analysis literature (see \cite{Maltoni:2004ei} for an example).

\item  I use $\chi^2_\Phi$ to find the best  value of $\Phi$, defined as the one that minimizes $\chi^2_\Phi$, and the 99\% C.L. interval for it, defined by the difference $\chi^2_\Phi - \chi^2_{\Phi, min}=6.65$.
 
\end{itemize}

Results are given for three cuts in
the neutrino energy: $E>19.3, 11.3, 5.3$ MeV.  The first
corresponds to the threshold applied in the search of the \df\ at SK
\cite{Malek:2002ns}, while the second represents a potential
improvement that SK could achieve with the proposed Gadolinium addition
\cite{Beacom:2003nk}.  The last cut, $E>5.3$ MeV, is the technical
limit given by the photomultipliers distribution in the SK tank and
would apply only in the -- currently unfeasible -- case of complete
background subtraction.   It  should be stressed that with such low  threshold the  \df\ starts to have a non-negligible contribution (about $\sim 50\%$ \cite{Ando:2004hc}) from supernovae at $z>1$, and thus it depends on the value of the power $\alpha$ in Eq. (\ref{eq:powerlaw}).    Here I used $\alpha=0$, therefore the results of this work for the lowest threshold have indicative character only; they could change by several tens of per cent for different values of $\alpha$.

 The effects of neutrino flavor conversion inside the star are included.
I calculated the effects of oscillations inside the Earth,  
using a realistic matter
density profile \cite{prem}, and in the assumption of isotropic \df. I find that these oscillations affect the \df\ by less  than $\sim 10\%$, and therefore I neglect them for simplicity.  

The same procedure of marginalization discussed for the flux can be done for the number of events in a detector. Here I do the calculation for SK and the 
inverse beta decay events: $\barnue + p \rightarrow n + e^+$.  This is by far the dominant signal in water, which justifies neglecting other reactions.  
I took the fiducial volume of SK of 22.5 kilotons and detection efficiency of  $\simeq 93\%$ above $\sim 7$ MeV  \cite{Hirata:1987hu,Jegerlehner:1996kx}.    In reality, the efficiency depends on the specific experimental cuts.  In the case of the published SK analysis the efficiency is 47\% (79\%) below (above) 34 MeV \cite{malekthesis}.  Thus, considering the exponential decay of the flux with the energy, the rates given here would have to be rescaled down by roughly a factor of 2 to be applicable to the current SK setup.  

\subsection{Results}

\begin{figure}[htbp]
  \centering
\includegraphics[width=0.9 \textwidth]{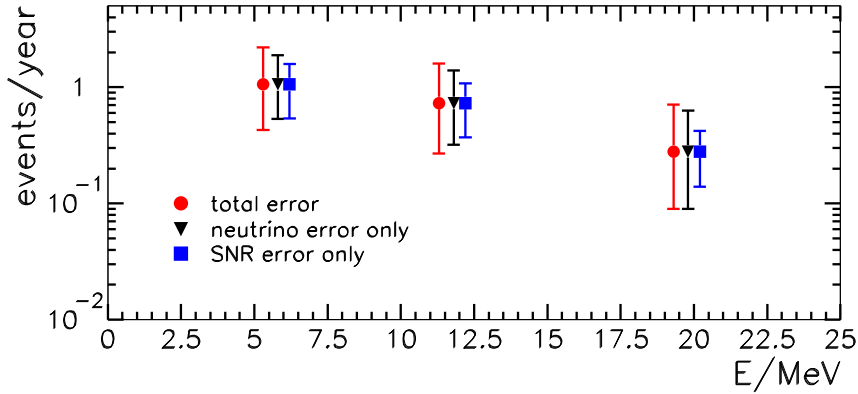}
\caption{The number of $\barnue + p \rightarrow n + e^+$ events per year
at SuperKamiokande as a function of the energy threshold $E_{th}$ (horizontal axis).  Results are given for $E_{th}=5.3, 11.3, 19.3$ MeV in neutrino energy. For each of these, three cases are shown (shifted in energy for visual convenience): (i) total error, from the marginalization of the likelihood function in the whole parameter space, (ii) error from the neutrino spectrum only, obtained by marginalization with the supernova rate function fixed at its best fit point, (iii) error from the supernova rate only, obtained by keeping the neutrino parameters fixed at the best fit values. The markers give the values in the best fit point, while the bars represent the 99\% C.L. interval (``error").  }
\label{number_of_events}
\end{figure}

The results for the flux and numbers of events are presented in Table \ref{summarytable} and Fig. \ref{number_of_events}. 

For the current SK threshold of $19.3$ MeV, the $\barnue$  flux in a detector is of the order of $\sim 10^{-1}~{\rm cm^{-2} s^{-1}}$. The value of maximum likelihood is $\Phi= 0.15~{\rm cm^{-2} s^{-1}}$ and is obtained with the parameters that minimize both $\chi^2_{87}$ and $\chi^2_{SNR}$, Secs. \ref{87} and \ref{snr}.    With 99\% C.L., the flux  must be larger than $ \sim 0.05~{\rm cm^{-2} s^{-1}}$ and can not exceed  $ 0.35~{\rm cm^{-2} s^{-1}}$, almost a factor of 4 below the SK limit.   The event rate is below 0.7 events/year.  The width of the 99\% interval is about a factor of 7-8 both in the flux and event rate.  

The flux  increases by almost one order of magnitude if the threshold is lowered to 11.3 MeV. This is explained by the exponential decay of the flux with energy.  The event rate instead increases more moderately with the lowering of the threshold.
This is because the detection cross section is proportional to the square of the energy, $\sigma \propto E^2$, and therefore the spectrum of the observed positrons decays less rapidly than the neutrino spectrum.  From Table \ref{summarytable} it appears that  for threshold at 11.3 MeV or lower the goal of one event/year is possible, but not guaranteed.

To expect at least one event/year with 99\% C.L.  a larger detector than SK is necessary. A volume of water  2.5 times the volume of SK would be sufficient for the lowest threshold of 5.3 MeV, while  with the current SK threshold a detector $\sim 10$ times larger is required.  This may become a reality with the next generation Megaton detectors.   With a fiducial volume 20 times larger than SK,  these detectors would register  $\sim 2 - 14 $ events/year for the higher threshold, with a maximum of $\sim 44$ events/year with detection above 5.3 MeV.

\begin{table*}
\centering
\begin{tabular}{|l|l|l|l|l|}
\hline
\hline
  & $E>19.3~{\rm MeV}$ & $E>11.3~{\rm MeV}$ & $E>5.3~{\rm MeV}$ \\
  & $\Phi/({\rm cm^{-2} s^{-1}})$   & $\Phi/({\rm cm^{-2} s^{-1}})$  & $\Phi/({\rm cm^{-2} s^{-1}})$  \\
 &  $\left[ N/{\rm yr^{-1}} \right]$  &  $\left[ N/{\rm yr^{-1}}\right]$ & $\left[ N/{\rm yr^{-1}}\right]$ \\
\hline
\hline 
best  &  0.15 &  0.93  &  9.4   \\
  &  $\left[  0.28 \right]$ &  $\left[ 0.73  \right]$ &    $\left[ 1.1  \right]$ \\
  \hline
 68\% C.L. &  0.12 - 0.21  &  0.64 - 1.19  &    6.4  - 13.5   \\
          &  $\left[ 0.19 - 0.41  \right]$  & $\left[ 0.51-0.85  \right]$ &  $\left[0.77 - 1.43\right]$\\
\hline
 90\% C.L. &  0.08 - 0.27   &   0.50 - 1.39 &   4.9 - 16.7   \\
          &  $\left[ 0.14-  0.52 \right]$  & $\left[ 0.40- 1.23 \right]$ &  $\left[ 0.62 - 1.72\right]$\\
\hline
 99\% C.L. &  0.05 - 0.35  &  0.33 - 2.1 &   3.2 - 22.6  \\
          &  $\left[  0.09 - 0.7 \right]$  & $\left[ 0.27 - 1.6 \right]$ &  $\left[ 0.43 - 2.2  \right]$\\
\hline
\hline
\end{tabular}
\caption{The predicted flux of $\barnue$ in a detector in the point of maximum likelihood and in the intervals of 68, 90, 99\% C.L.
for different energy
thresholds.  The numbers in brackets have the same meaning, but for the rates of $\barnue + p \rightarrow n + e^+$ events at SuperKamiokande for  $\sim 93\%$ detection efficiency (see text). } 
\label{summarytable}
\end{table*}

What is the contribution of the individual sources of error to the total uncertainty on the \df or event rate? Fig. \ref{number_of_events} answers this question for the event rate. For the three thresholds, the figure shows the error bars obtained by marginalizing the likelihood over a subset of parameters while keeping the others fixed at their best fit values.  It appears that the error bar from the \snr\ (with fixed neutrino parameters) is smaller than the one due to the neutrino spectra (e.g. fixed \snr).   The first is about a factor of 3 wide, while for the second the width  is  a factor of 4-7 depending on the threshold.  Notice that this width is larger for higher threshold, reflecting the larger uncertainty in the high energy tail of the SN1987A data with respect to their full spectrum. 

Let me now discuss the validity of the results and possible generalizations.
A relevant question is how the interval of values of the \df\
depends on the details of the analysis of the SN1987A data, and, in
particular, on possible priors on the neutrino parameters.  I checked
that the results have little dependence on those.  For example, the
interval for the flux above 19.3 MeV (see Table \ref{summarytable}) is
unchanged if I impose (simultaneously) the constraints: $10~{\rm
MeV}<E_{0 \bar e }<20~{\rm MeV}$, $E_{0 \bar e }<E_{x}$ and
$0.5<L_{\bar e }/L_{x}<2$.  This is explained with the parameter space
being largely degenerate, so that a restricted portion of it still
covers almost all the physically different possibilities.  For the
same reasons the results are also practically insensitive to different
parameterizations of the original fluxes (e.g., different
$\alpha_{\bar e}$ and/or $\alpha_{x}$)\footnote{Clearly, different
parameterizations of the original fluxes give differences in the
region of the energies and luminosities allowed by SN1987A, however.}.

The analysis here does not include systematic errors.  Still, here I briefly discuss  how the prediction of the \df\ would change if some of the inputs turned out to be systematically wrong and were corrected.   
Let us consider the event in which  $R_{SN}(0)$ is higher than what used here (see discussion in Sec. \ref{snr}).  With $R_{SN}(0)$ increased by a factor of two (three) the maximum likelihood value of the flux is $\Phi=0.20~{\rm cm^{-2} s^{-1}}$ ($\Phi=0.23~{\rm cm^{-2} s^{-1}}$)  above the current SK threshold of $19.3$ MeV.  This is at most a $\sim 50\%$ increase with respect to the result in Table \ref{summarytable}. 
Another example refers to the poorly known normalization of the \snr.  Specifically, my results could become a factor of two higher  if it is confirmed that normalization of the \snr\ is two times higher as favored by the \sfr\ measurements (Sec. \ref{snr} and Fig. \ref{snrband}). A combined fit of the supernova data and \sfr\ measurements would reduce the uncertainty on the slope $\beta$ and therefore the total error on the \df.  To check how this error would change, I performed the same calculation outlined in this section with $\beta$ fixed at its best fit value, $\beta=3.44 $, which happens to be nearly the same both in the \snr\  and in the \sfr\ data analyses (Sec. \ref{snr}).  I find that the results for the \df\ do not change appreciably, due to the large degeneracies between different sets of parameters. 

Besides technical details, it has to be stressed that the results
given here are valid if the SN1987A neutrino flux is typical and thus
represents a generic output of a core collapse supernova.  This
question will be answered by data from a future galactic
supernova. At the moment, recent calculations show that the main observational
features of SN1987A are reproduced by the standard neutrino-driven
explosion mechanism, disfavoring possible anomalies in the SN1987A
event \cite{Kifonidis:2005yj}.


\section{Discussion and conclusions}
\label{concl}

Let me summarize this work.  I have calculated the $\barnue$ component
of the \df\ in a detector, using the information on the neutrino
spectra from SN1987A and the information on the supernova rate 
from direct observations of core collapse supernovae.  I calculated  the likelihood functions for the SN1987A data and for the supernova rate measurements,  and marginalized the combined likelihood to find the interval of \df\ allowed at a given confidence level.

The SN1987A data favor a composite neutrino spectrum, meaning that a
scenario with inverted mass hierarchy and $\sin^2 \theta_{13}\gta
10^{-4} $ is disfavored (in agreement with refs. \cite{Lunardini:2000sw,Minakata:2000rx}).  The best
fit spectrum reproduces both the K2 and IMB data well, even if IMB alone
favors a thermal spectrum over a composite one. It is characterized by
a very soft and very luminous component, with parameters in contrast
with general theoretical arguments and with numerical simulations.
The allowed region of the parameter space, however, is very extended
and includes more natural scenarios at 68\% C.L.. 

The supernova rate measurements allow a present rate of about
$(0.2 - 1.2) \cdot 10^{-4}~{\rm Mpc^{-3}~yr^{-1}}$ and a power of increase with $z$ of
about 2 - 6 .  Also in this case a large region of parameters is allowed.
The function that best fits the data has similar power, $\beta \sim 3$, but
different normalization (about a factor of two smaller) with respect
to what is inferred using data on the star formation rate and the
proportionality of this rate to the supernova rate.

The results for the \df\ ($\barnue$ component) show that the flux
above the current SK threshold is likely to be about one order of
magnitude below the current upper limit of $1.2 ~{\rm cm^{-2} s^{-1}}$, and is smaller than this limit by a
factor of $ \sim 4$ with 99\% confidence.  This factor of 4 is the minimal
improvement that SK should achieve if the energy threshold remains the
same. Any lowering of the threshold would significantly enhance the
possibility to detect the \df. Still, however, if my prediction is
correct, a Megaton detector would be necessary to expect a detection
with very high confidence. 

\begin{figure}[htbp]
  \centering
  \includegraphics[width=0.990\textwidth]{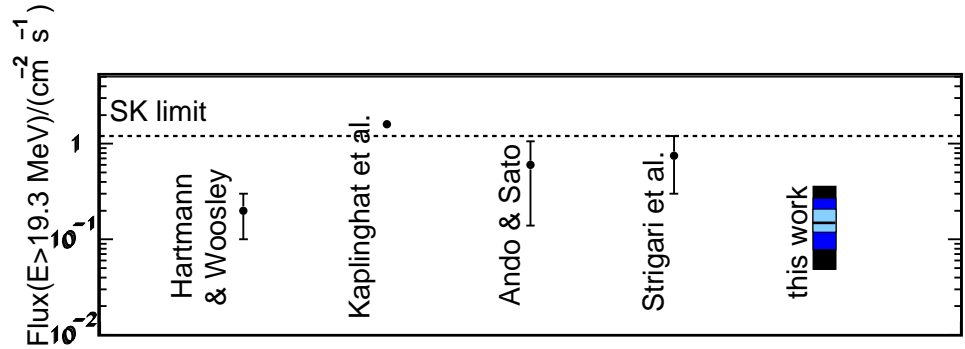}
\caption{The results of different authors (from left to right: refs. \cite{Hartmann:1997qe}, \cite{Kaplinghat:1999xi} (upper bound only), \cite{Ando:2004sb,Ando:2004hc}, \cite{Strigari:2003ig} and this work) for the flux of $\barnue$ above a threshold of 19.3 MeV (in neutrino energy). For this work the colored (shaded) bands  correspond to 68\%, 90\% and 99\% confidence levels, with the central like marking the maximum likelihood value. The SK limit is shown as well (dashed line).} 
\label{compare}
\end{figure}

How does this work compare to others in the field?  Fig. \ref{compare}
answers this question, by presenting the results of different papers
\cite{Hartmann:1997qe,Kaplinghat:1999xi,Ando:2004sb,Ando:2004hc,Strigari:2003ig}
for the $\barnue$ diffuse flux above $19.3$ MeV, with the interval of
uncertainty associated to it. The authors quoted in the figure are
only a small sample of all those that have calculated the \df; they
were chosen for illustration purpose and because their results can be
directly compared to each other's and to mine.  Each of them used
neutrino spectra from a number of numerical simulations, and inferred
the supernova rate from the star formation rate.  They also gave a
tentative uncertainty on their result, associated with the lack of
consensus on numerical codes and with the error on the star formation
rate.  The error bars in the figure, as well as the central points, do
not have a statistical meaning except for the result of this paper,
where the shaded areas represent the intervals allowed at 68\%, 90\%
and 99\% C.L. and the central line marks the maximum likelihood value.

From the figure it appears that the interval of flux calculated here
 overlaps with the prediction of other authors within the error, but
 extends to lower values. In particular, it is the only calculation
 which allows a flux smaller than $\sim 0.1~{\rm cm^{-2} s^{-1}}$.
 The differences between my results and the larger flux -- close to
 the SK limit -- allowed by Strigari et al. \cite{Strigari:2003ig}
 and by Ando and Sato \cite{Ando:2004sb,Ando:2004hc}, are due to the
 different inputs used by these authors: a factor of about two higher
 \snr\ (see discussion in Sec. \ref{snr}) and larger neutrino flux at
 high energy, due to the harder neutrino spectra used, with $E_{0x}$
 up to $\sim 24$ MeV. These spectra are motivated by numerical
 calculations of neutrino transport, but are known to be at best
 marginally compatible with the SN1987A data
 \cite{Jegerlehner:1996kx,Minakata:2000rx,Kachelriess:2000fe}, a fact
 that this paper confirms\footnote{This poor or no compatibility is
 not surprising. Indeed, numerical calculations are currently limited
 to rather crude neutrino transport or to simulating only the first
 second or so after the core bounce. Therefore, they are not directly
 comparable to the SN1987A data. }.  Hartmann and Woosley
 \cite{Hartmann:1997qe} used a supernova rate function similar to that
 of Strigari et al. and of Ando and Sato, together with softer
 neutrino spectrum: a thermal spectrum with temperature of $\sim $ 4
 MeV.  This explains their finding a lower flux with respect to other
 authors.  
The result by Kaplinghat et al. \cite{Kaplinghat:1999xi} is a conservative upper bound obtained with a unoscillated  neutrino spectrum and the maximum \snr\ compatible with metal enrichment history. 

The comparison with the previous literature shows that this
work is complementary to others.  It is especially useful
because it gives an idea of how the predicted \df\ can change with a
change of the method of calculation or of the inputs.  
This should be encouraging  to work to improve our knowledge of the inputs, in particular of the neutrino spectra and of the normalization of the \snr.
This paper also makes the point that the flux may be smaller than generally expected, thus supporting the case for a next generation of neutrino detectors with larger volumes and/or lower energy thresholds.

\ack
I am very grateful to A. Friedland, W. Haxton and A. Yu. Smirnov for
helpful discussions and comments on the first version of the manuscript. I warmly thank E. Cappellaro, M. Giavalisco,  M. Maltoni and M. Turatto for fruitful private communications.  I am indebted to the anonymous referee for crucial  comments. I acknowledge
support from the INT-SCiDAC grant number DE-FC02-01ER41187.


\bibliographystyle{apsrev}


\end{document}